\documentclass[conference]{IEEEtran}
\IEEEoverridecommandlockouts
\usepackage{cite}
\usepackage{amsmath,amssymb,amsfonts}
\usepackage{algorithmic}
\usepackage{graphicx}
\usepackage{textcomp}
\usepackage{xcolor}
\usepackage{url}
\usepackage{float} 
\usepackage{booktabs}

\def\BibTeX{{\rm B\kern-.05em{\sc i\kern-.025em b}\kern-.08em
    T\kern-.1667em\lower.7ex\hbox{E}\kern-.125emX}}
    
\begin{document}

\title{Using ML-based Regression Techniques to Mitigate GOES Energetic Proton Flux Data Contamination and Magnetospheric Effects}

\author{ \IEEEauthorblockN{Aatiya Ali}
    \IEEEauthorblockA{\textit{Department of Physics and Astronomy} \\
    \textit{Georgia State University}, \\Atlanta, USA\\0000-0002-0786-7307}
    \and
    \IEEEauthorblockN{Viacheslav Sadykov}
    \IEEEauthorblockA{\textit{Department of Physics and Astronomy}\\
    \textit{Georgia State University}, \\Atlanta, USA\\0000-0002-4001-1295}}

\maketitle

\begin{abstract}
Positioned at geostationary orbit (GEO) $\sim$36,000 km above Earth, NOAA's GOES series has recorded real-time energetic proton flux measurements crucial for space weather monitoring for over three decades. Although machine learning models have advanced solar energetic particle (SEP) event prediction using GOES data, the sudden yet sparse nature of SEP events necessitates high-quality proton flux measurements. Previous studies have identified contamination issues in GOES data, when the presence of higher-energy protons can cause parasitic signals in lower-energy GOES channels and lead to artificially elevated fluxes in lower energy ranges (e.g., 10 - 50 MeV, \cite{contamination_noaa,goes_contamination_posner,rodriguez_cont}). As of now, no universal correction method has been implemented for the publicly available NOAA data. In addition, the effects of Earth's magnetosphere on the 10 - 50\,MeV particles are not fully understood yet. This study assesses a reconstruction method using concurrent solar proton event (SPE) measurements from SOHO-EPHIN, which align well with GOES measurements of SPEs across solar cycles 23 and the bulk of cycle 24, but represent the off-magnetospheric environment of the Lagrange 1 point. We train regression models on GOES proton fluxes across multiple energy bins, employing EPHIN fluxes as prediction targets. We expect that similar approaches can allow us to derive non-contaminated flux proxies that preserve valuable data and more accurately capture the characteristics of SPEs, providing a more stable dataset for analyzing SEP behavior and potentially improving SEP event prediction models.
\end{abstract}


\section{Introduction}
A hazardous subclass of solar energetic particle (SEP) events, solar proton events (SPEs), is traditionally defined as instances where protons with energies $\geq$10 MeV exceed a flux of 10 particle flux units (pfu) \cite{sep_energies_anastasiadis}. The high-energy protons in these events can increase radiation levels, posing health risks to astronauts, interfering with electronic systems, and jeopardizing both crew safety and mission success. They can also cause radiation-induced damage or system failures in equipment and satellites \cite{health_risk_onorato,sickness_lee}. With growing relevance for space exploration and commercial space travel, it is imperative to develop reliable methods to predict these SPEs to mitigate their effects. In particular, lunar exploration is uniquely affected by SEPs. Positioned at an average distance of $\sim$380,000 km from Earth, the Moon remains outside Earth’s magnetosphere for most of its orbit, exposing its surface to radiation from solar wind ions, high-energy SEPs, and galactic cosmic rays (GCRs) with minimal attenuation \cite{radiation_dandouras}. In contrast, on Earth, the magnetosphere can act as a shield, scattering low-energy particles back into space and limiting their access to the geostationary and low-Earth orbits \cite{bsphere_deflecion_liu}. Therefore, it becomes essential to understand the commonalities and differences between SEP events observed in two distinct environments: those without a protective magnetosphere (such as the Moon or the Lagrange point 1 (L1)) and those within Earth's magnetosphere. In addition to the point above, there exist instrumental differences between GOES and EPHIN that also affect their measurements. For example, during intense SEP events, lower-energy channels of GOES instruments can sometimes record falsely elevated signals due to contamination from high-energy particles, such as relativistic electrons or protons in the upper MeV to GeV range \cite{goes_contamination_posner}. The comparison of these observations and the mitigation of instrumental effects is therefore critical for predicting astronaut exposure risks, assessing long-term effects of space weather on the lunar environment, and understanding the mechanisms by which SEPs reach the Moon \cite{main_moon_liuzzo}.  

\subsection{Scope of this Work}
This study extends earlier analyses of proton flux at geostationary orbit (GEO) during Solar Cycle 23 and much of Cycle 24 by comparing those SPE characteristics with simultaneous measurements from EPHIN at L1, located beyond Earth’s magnetosphere \cite{me1}. While not exactly at the location of the Moon, EPHIN measurements more closely represent the lunar environment than GOES data, as they sample SEPs beyond Earth’s magnetosphere while remaining in near-Earth space. This work specifically focuses on 10 - 50\,MeV proton fluxes. Although the 10 – 50\,MeV proton range does not directly represent the integrated $\geq$10 MeV flux used to define NOAA’s solar radiation storm scales (S-scales)\footnote{\url{https://www.swpc.noaa.gov/noaa-scales-explanation}}, it typically accounts for the bulk of the proton flux during weaker SEP events and dominates the peak fluxes in stronger events. By comparing SPE properties observed at the L1 point with those detected at GEO, {one can} identify patterns or discrepancies between their statistics and explore how various factors may influence changes in fluxes of energetic protons as they travel from L1 to GEO. The key elements required for this comparison are summarized in this paper, and a more detailed analysis is performed in \cite{sep_comparison_ali}. Exploring these dynamics could provide valuable insights into particle precipitation into the Earth's magnetosphere and, therefore, the refinement of SPE prediction models for lunar operations. 
\begin{figure*}[htbp!]
\centering 
\includegraphics[width=0.8\textwidth]{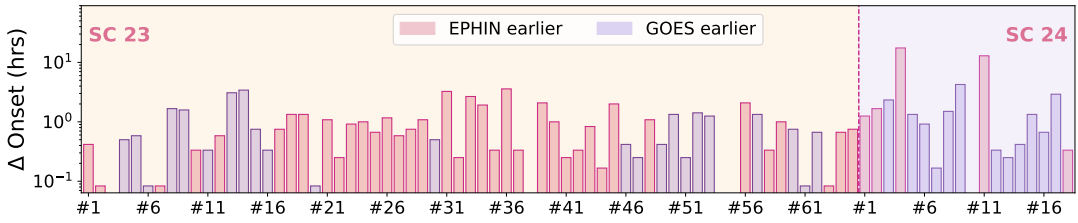} 
   \caption{Differences in SPE start times observed by GOES and EPHIN across SCs 23 \& 24. Each bar represents a single SPE, with the x-axis enumerating the events. Bar colors indicate which instrument detected the earlier time, while the background reflects the corresponding SC. Instances where bars are absent (e.g., onset for event \#3) indicate no measurable offset, i.e., both instruments detected the property simultaneously ($\Delta$ = 0 hrs).} \label{starts}
\end{figure*}

During a separate study comparing SPE characteristics at L1 and GEO, we identified unexpected trends that motivated the development of this work, which applies machine learning–based regression techniques to mitigate possible flux contamination in GOES data \cite{sep_comparison_ali}. 

As shown in Fig. \ref{starts} (reproduced from \cite{sep_comparison_ali}), GOES occasionally identified 10-50\,MeV $\geq$10\,pfu SPE onsets earlier than SOHO/EPHIN during SC 23, but did so almost consistently in SC 24, even though GEO is positioned farther from the Sun than L1. While 10 MeV protons can theoretically travel from L1 to Earth in roughly one minute, onset delays in some events extend to several hours. Notably, we observe pronounced differences in peak flux measurements, with GOES reporting significantly higher values during several events. These discrepancies may arise from localized enhancements, preferential transport effects, or contamination in the 10 – 50 MeV channel due to parasitic signals from higher-energy protons \cite{transport_effects_battarbee}. This is evident in Fig. \ref{fig:ratio} (reproduced from \cite{sep_comparison_ali}), where SPE peak fluxes and fluences generally follow the one-to-one relationship (dashed magenta line) between instruments, but diverge for the most intense events- the subset in the upper-right region of both plots. In these cases, GOES records orders of magnitude more counts than EPHIN, suggesting that during extreme events, its 10 – 50 MeV measurements may be inflated by parasitic signals from high-energy passbands. Such systematic overestimation affects not only peak fluxes but also integrated quantities such as fluences. \textit{Together, these results underscore how contamination in GOES proton channels can lead to mischaracterization of SPE intensities, which are often used as targets in forecasting models. Correcting these biases would substantially improve forecast reliability, particularly for near-Earth and lunar mission planning.}
\section{GOES Proton Flux Data}
\begin{figure}[H]
\centering 
 \includegraphics[width=.45\textwidth]{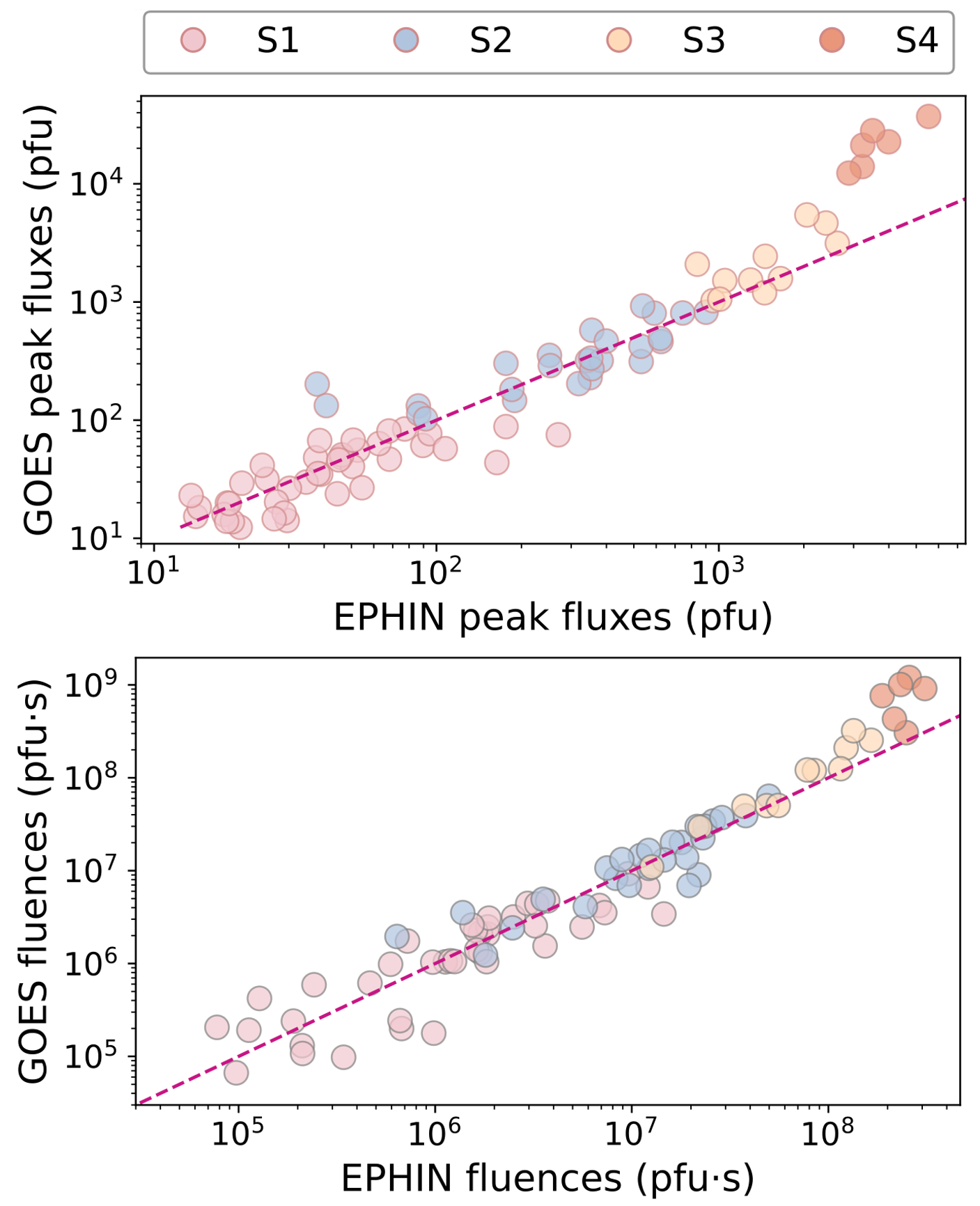}
    \caption{Comparison of GOES peak fluxes (left) and fluences (right) with EPHIN during SPEs across SCs 23 \& 24.}
  \label{fig:ratio}
\end{figure}

The GOES energetic proton flux data is valued for its coverage of SEP events and is widely used for scientific and space weather applications \cite{pl_example_aminalragia}. Our previously developed SPE-detection algorithm used $\geq$10 MeV proton flux data from NOAA's GOES satellites (GOES-08 to GOES-15) spanning SCs 22-24 \cite{me1}. With the launch of GOES-13 in 2011, the Energetic Proton, Electron, and Alpha Detector (EPEAD) replaced the Energetic Particle Sensor (EPS), still providing similar integrated SEP data products. Unlike the EPS, which was single-direction-oriented, the updated EPEAD now features detectors facing both East and West. To ensure data consistency, NOAA designates a ``primary'' satellite during overlapping data recording periods. Similarly, we select a ``primary'' instrument for each month based on an empirical approach, giving priority to either the East or West detector, depending on which measures the highest peak proton flux during the SPEs of that month. GOES typically reports integral proton flux products across various energy bins: $\geq$1 MeV, $\geq$5 MeV, $\geq$10 MeV, $\geq$30 MeV, $\geq$50 MeV, $\geq$60 MeV, and $\geq$100 MeV. These fluxes are computed from the differential flux measurements, assuming the conditions described below. As a result, the fluxes recorded for the $\geq$1 MeV integral include contributions from all higher energy channels, including those above 100\,MeV. While most of the GOES data are corrected for galactic cosmic background and high-energy particles penetrating the shielding, NOAA warns about potential cross-contamination due to contamination from high-energy particles, such as relativistic electrons or protons in the upper MeV to GeV range \cite{goes_contamination_posner}. The GOES Energetic Particle Correction Algorithm calculates differential and integral proton fluxes by first determining the background count rate for each energy channel. A filter technique is applied to update the background estimate over time, and this value is subsequently subtracted from the measured particle counts. The algorithm then assumes that the SEP energy spectrum between \textit{successive} energy channels can be approximated by a simple power-law distribution. These corrections are applied independently to each data segment, enabling continuous background-corrected flux estimation for operational use. However, this assumption may not hold during the early phases of SPEs, when velocity dispersion temporarily causes the proton spectrum at 1 AU to exhibit intensities increasing with energy, disrupting the flux correction process \cite{goes_contamination_posner}. Rodriguez \textit{et al.} \cite{contamination_noaa} address this issue by detailing the background and proton contamination corrections applied exclusively to the averaged 1-minute cadence flux data. However, these corrections are not applied to the 5-minute cadence data used in real-time processing and in this study. Overall, contamination likely accounts for the counterintuitive patterns seen in Figs. \ref{starts} and \ref{fig:ratio}, where GOES appears to register SPE onsets earlier than EPHIN in several cases and records substantially higher peak fluxes and total particle counts- sometimes by orders of magnitude. High-energy particles (e.g., $>$100 MeV protons) travel faster and reach GOES detectors earlier than protons in the 10 – 50 MeV range, which is the focus of this analysis. When fluxes detected in higher-energy channels spill over into lower-energy channels, the $\geq$10 pfu SPE threshold can be reached prematurely, triggering the start of an event earlier than it should. The same issue can affect the recorded peak flux times of events. The strongest events are associated with higher fluence measurements recorded by GOES, but these measurements most likely include artificially elevated particle counts from high-energy particle contamination, leading to an overestimation of event intensity.  
 \vspace{0cm}
\section{Characterization of GOES Proton Flux Contamination}
Following NOAA’s S-scale classification, which categorizes events from S1 (minor) to S5 (extreme) based on the fluence of $\geq$10 MeV protons, we identify SPEs as intervals where proton fluxes in the 10 – 50 MeV range exceed 10 pfu for at least 15 minutes (three consecutive data points in both instruments). For GOES data, we utilize the publicly available 5-minute averaged integrated corrected proton flux measurements provided by NOAA's National Centers for Environmental Information (NCEI)\footnote{\url{https://www.ncei.noaa.gov/data/goes-space-environment-monitor/access/}}. The $\geq$50\,MeV flux is subtracted from the $\geq$10\,MeV flux to generate the 10 - 50 MeV proton flux product. For the SOHO/EPHIN proton flux measurements, we utilize the data obtained using the Relativistic Electron Alert System for Exploration (REleASE) algorithm by \cite{goes_contamination_posner} accessible via Zenodo\footnote{\url{https://zenodo.org/records/14191918}}. The fluxes in energy bins from \#15 to \#28 are summed to generate the 10 - 50 MeV proton flux product. The time period from 1995 to 2016 is considered for analysis. Events are identified independently for GOES and EPHIN, after which we compile a catalog of SPEs detected concurrently at both locations, yielding a total of 83 events.

For each event, we document properties including start and end times, peak flux and its timing, and event fluence as measured by each instrument. To avoid overcounting due to flux oscillations near the 10 pfu threshold, events separated by less than 10 minutes are merged, as these are likely to represent minor variations within a single event rather than distinct SPEs. For our convenience, we refer to SPEs that potentially experience high-energy particle contamination as ``contamination candidates (CCs).'' We note here that we do not confirm whether specific SPE data are truly contaminated in this study, but we want to explore how our GOES-to-EPHIN regression performs on these presumably contaminated data, as well as make sure that there were events most likely experiencing contamination in each of the data partitions used for training, validation, and testing. We use the following criteria based on the integral fluxes to identify non-exclusively SPEs that may potentially have contaminated GOES flux data, hereafter referred to as CCs:
\begin{figure*}[ht!]
\centering 
 \includegraphics[width=.7\textwidth]{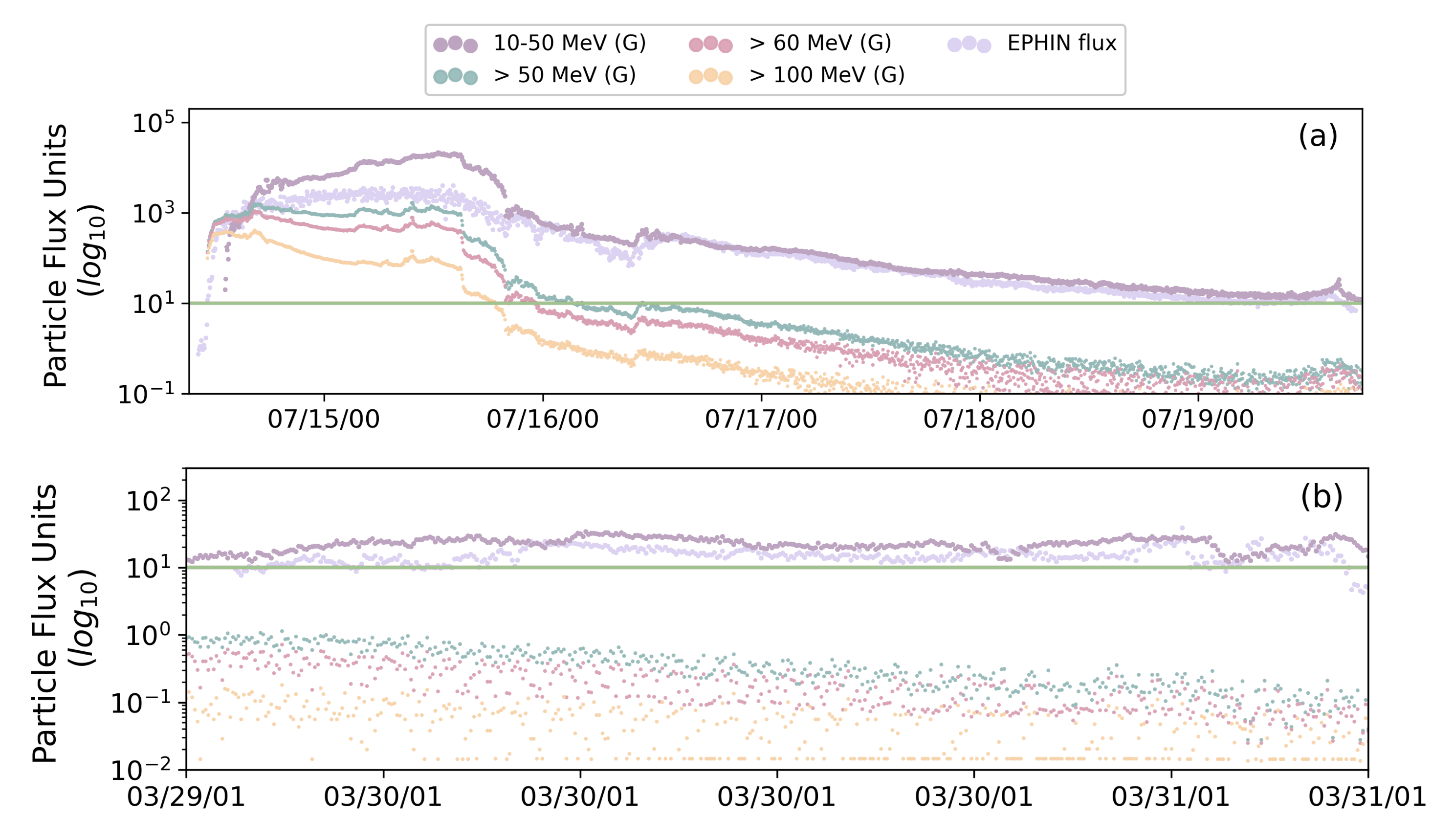}
  \caption{(a) An SPE where the 10 – 50 MeV proton flux profile closely mirrors the signals in the higher-energy GOES channels, indicating potential contamination. (b) An SPE showing no obvious contamination signatures, with proton channels above 50 MeV showing significantly lower signals compared to the 10 – 50 MeV channel, as expected.}
  \label{fig:issue-and-fix}
\end{figure*}
  
\begin{itemize} 

\item If increases in the 10 - 50 MeV fluxes occur simultaneously with those in the higher-energy channels ($>$50 MeV, $>$60 MeV, $>$100 MeV) and if there is no CME shock arrival at L1 at around that time, this raises contamination concerns. Since 10 - 50 MeV protons should take longer to reach the detector, such early increases suggest contamination rather than realistic physical changes.
\item Concurrently with the point above, rapid, unique changes in the flux of higher-energy channels should not be mirrored in the lower-energy channels unless contamination is present.
\item If GOES fluxes are notably larger than EPHIN fluxes, it suggests a high likelihood of data contamination. Since EPHIN is not subject to the same contamination issues as GOES, any significant or prolonged difference between GOES and EPHIN measurements during such episodes should be weak or negative. Large discrepancies favoring GOES indicate that its data may be affected. 
\end{itemize}

Using these criteria, we examined the 83 SPEs in our sample individually and designated them as either CCs (17 SPEs) or non-CCs (66 SPEs). Figure \ref{fig:issue-and-fix} compares these scenarios: (top) contamination of the 10 - 50 MeV channel by higher-energy protons, and (bottom) an event without the obvious signs of contamination in integral channels. In the top panel, the 10 - 50 MeV measurements exceed the 10 pfu threshold only 10 minutes after the $\geq$50 MeV, $\geq$60 MeV, and $\geq$100 MeV channels surpass the same threshold. This short delay suggests that the 10 - 50 MeV measurements may be affected by contamination from higher-energy channels, as a longer delay would be expected if the lower-energy measurements were unaffected. Additionally, multiple abrupt changes in the higher-energy flux profiles are mirrored in the 10 - 50 MeV profile, further supporting the idea that stronger fluxes influence the 10 - 50 MeV measurements. The significant and prolonged difference in fluxes between EPHIN and GOES during the first day of the SPE is also unexpected and raises concerns, as it is not entirely realistic given the positioning of both instruments. Taken together, these factors lead us to classify the SPE in the top panel as a CC. In contrast, the bottom panel does not exhibit these issues and is classified as a non-CC. While these questionable fluxes may reflect actual SEP behavior, they also raise the possibility of additional undetected contamination in GOES data, and we note here that the additional analysis of differential fluxes may be required to detect truly contaminated events. These challenges in reliably utilizing GOES proton flux data highlight the importance of evaluating and accounting for SEP contamination, particularly during intense SEP events. In this work, we aim to develop methods to correct the affected data and improve their reliability.

In addition, SEP measurements at GEO by GOES may be influenced by magnetospheric shielding and modulation. Such effects can inhibit particle access and extend the observed decay phase of events. O'Brien \textit{\textit{et al.} }\cite{particle_energy_obrien} emphasized the role of particle energy in determining SEP access to different regions of Earth’s magnetosphere. Other studies have examined how various magnetospheric processes influence SEP propagation and the particle populations that reach GEO. For example, Filwett \textit{et al.} \cite{arrival_direction_filwett} and Kress \textit{et al.} \cite{kress_cutoff} demonstrate how geomagnetic indices such as \textit{AE, Kp, Ap,} and \textit{Dst} play a critical role in driving magnetospheric cutoff boundaries, thereby shaping the particle populations able to penetrate to GEO- particularly during strong SPEs. To investigate similar effects, we analyze hourly averaged geomagnetic indices from OMNI2 data\footnote{\url{https://spdf.gsfc.nasa.gov/pub/data/omni/low_res_omni/}} obtained from spacecraft near Earth and at L1, and include them as inputs into the regression models along with the energetic proton fluxes.

\section{Machine Learning–Based Approaches to GOES-EPHIN Flux Regression}
Regression models are a well-suited approach for the task of GOES-to-EPHIN flux correction because they can learn the relationship between various input features (in our case, GOES proton flux data from energy channels: 10 - 50 MeV, 50 - 60 MeV, 60 - 100 MeV, $>$100 MeV, and various geomagnetic indices) and the discrepancy introduced by contamination. Once trained, these models can infer corrections that help restore underlying physical signals. In our case, we use the simultaneously observed, non-contaminated fluxes from EPHIN as the reference targets during model training. These serve as proxies for the true proton fluxes, allowing the models to learn the mapping between contaminated GOES measurements and cleaner flux profiles. We also note that while contamination remains the primary issue, there could be additional magnetospheric effects to correct for. This allows the reduction of systematic noise and measurement biases in the GOES data by leveraging EPHIN's cleaner observations as a corrective benchmark. In doing this, we are able to quantify the degree of flux alteration and account for irregular, non-systematic instrumental contamination that may affect GOES proton measurements. In this study, we implement three machine learning (ML)-based regression models: Random Forests (RF), eXtreme Gradient Boosting (XGBoost\footnote{\url{https://xgboost.readthedocs.io/en/stable/}}), and Multi-Layer Perceptron (MLP) neural networks. 

RFs are ensemble-based bagging classifier methods that aggregate predictions from multiple decision trees with limited information propagating to each, offering strong performance on noisy data and enabling interpretable insights through feature importance rankings. XGBoost is a boosting tree-based ensemble technique and typically outperforms traditional decision trees by optimizing residual errors iteratively and reducing overfitting. Lastly, MLPs are powerful tools for learning smooth, nonlinear mappings between inputs and targets, making them suitable for modeling variations in flux profiles that may arise from data contamination. The use of regression techniques for flux prediction and correction in space weather has been successfully demonstrated in several recent studies. Stumpo \textit{\textit{et al.}} \cite{stumpo_rf} developed an RF model to forecast $>$10 MeV solar proton fluxes using relativistic electron measurements as inputs, achieving effective lead times and accurate flux predictions. Similarly, the Low Energy Electron MLT geosynchronous orbit Regression (LEEMYR) model developed by Miller \textit{\textit{et al.}} \cite{leemyr} employed regression techniques to forecast electron fluxes at GEO using GOES-16 observations, showing strong agreement with ground truth values and highlighting the applicability of data-driven correction methods in radiation belt contexts. Furthermore, solar wind speed forecasting studies have shown that gradient boosting regression models can outperform empirical and physics-based approaches such as Wang–Sheeley–Arge (WSA) by leveraging patterns from multi-source solar wind and coronal data, capturing nonlinear relationships and temporal dependencies that traditional models often fail to represent \cite{bailey}.

To evaluate the performance of the ML–based regression models, we train them using a combination of CC and non-CC GOES data to reflect realistic observational conditions. We note here that our current experiments are limited to modeling the concurrent fluxes and do not include any temporal shifts between GOES and EPHIN fluxes. The simultaneously measured EPHIN proton fluxes serve as the target outputs, providing a reference for the uncontaminated flux profiles. The goal is for the models to learn the underlying patterns in the GOES input data and reconstruct what the true proton fluxes would have been during periods affected by contamination. We assess model performance using the coefficient of determination ($R^2$) and mean squared error (MSE), which quantify how well the predicted flux values match the clean EPHIN reference data. The $R^2$ score quantifies the proportion of variability in the target data that the model can explain, with values approaching 1 indicating a stronger agreement between the predicted and true flux values. MSE measures the average squared difference between predicted and actual values, with lower values corresponding to more accurate predictions. Together, these metrics provide a robust evaluation of each model’s ability to reconstruct `clean' flux profiles from contaminated input data.

\texttt{GridSearch} is used to optimize model hyperparameters with respect to the ${R}^{2}$ score. The hyperparameters considered during \texttt{GridSearch} for each model are recorded in Table \ref{tab:grid}.

\begin{table}[htbp]
\centering 
\caption{Parameter grids considered during \texttt{GridSearch} for each model. Asterisk (*) denotes the most frequent optimal value.}
\begin{tabular}{p{0.13\linewidth} p{0.7\linewidth}}
\toprule
\textbf{Model} & \textbf{Parameter Grid} \\
\midrule
\textbf{MLP} & 
\texttt{hidden\_layer\_sizes}: [(150,), (128,64), (256,128,64), (200,100,50), (64,64), (50,25,10*)];  
\texttt{activation}: [``relu", ``tanh*", ``logistic"]; 
\texttt{solver}: [``adam", ``lbfgs*"]; 
\texttt{alpha}: [1e-8, 1e-7, 1e-6, 1e-5*, 1e-4, 1e-3, 1e-2]; 
\texttt{learning\_rate}: [``constant*", ``adaptive"]; 
\texttt{learning\_rate\_init}: [1e-4, 5e-4, 1e-3, 5e-3*]; 
\texttt{early\_stopping}: [False*]; 
\texttt{max\_iter}: [1500*, 3000, 5000]; 
\texttt{n\_iter\_no\_change}: [5, 10*, 20] \\
\textbf{RF} & 
\texttt{n\_estimators}: [400, 600*, 800, 1000, 1200]; 
\texttt{max\_depth}: [None, 20*, 30, 40]; 
\texttt{min\_samples\_split}: [2*, 4, 8, 16]; 
\texttt{min\_samples\_leaf}: [1*, 2, 4, 8]; 
\texttt{max\_leaf\_nodes}: [None*, 63, 127]; 
\texttt{max\_features}: [``sqrt*", 0.5, None] \\
\textbf{XGBoost} & 
\texttt{n\_estimators}: [200, 400, 800*, 1200]; 
\texttt{max\_depth}: [4, 6, 8*, 12]; 
\texttt{learning\_rate}: [0.005*, 0.01, 0.02, 0.05]; 
\texttt{subsample}: [0.5, 0.7*, 0.9, 1.0]; 
\texttt{min\_child\_weight}: [1, 3, 5*, 10]; 
\texttt{reg\_lambda}: [0.5, 1.0*, 2.0, 5.0]; 
\texttt{reg\_alpha}: [0.0*, 0.1, 0.5, 1.0]; 
\texttt{gamma}: [0.0*, 0.1, 0.5] \\
\bottomrule   
\end{tabular} \label{tab:grid}
\end{table}

For all models, input features are linearly interpolated between neighboring true values to address missing data, then paired with the corresponding EPHIN flux measurements as training targets. A two-fold cross-validation (CV) is employed during training to balance bias and variance, after which the model with the best parameters (found using \texttt{GridSearch)} is applied to the remaining event samples in the testing phase, and its predictions are compared to the EPHIN fluxes. This process is repeated across multiple independent runs to ensure robustness and reproducibility. To ensure a fair comparison between models, all sources of randomness are fixed during training and evaluation. Specifically, we set the NumPy and Python random seeds (\texttt{np.random.seed(42)} and \texttt{random.seed(42)}) and assigned \texttt{random\_state=42} to all relevant \texttt{scikit-learn} components. This guarantees that each model is trained, validated, and tested on identical data splits (event samples), enabling reproducible and unbiased performance comparisons. We also make sure that the data from any unique SPE is confined to a single partition (train, validation, or test) to avoid artificial correlations between data sets. 

\section{Results}
\begin{figure}[ht!]
\centering 
\includegraphics[width=.49\textwidth]{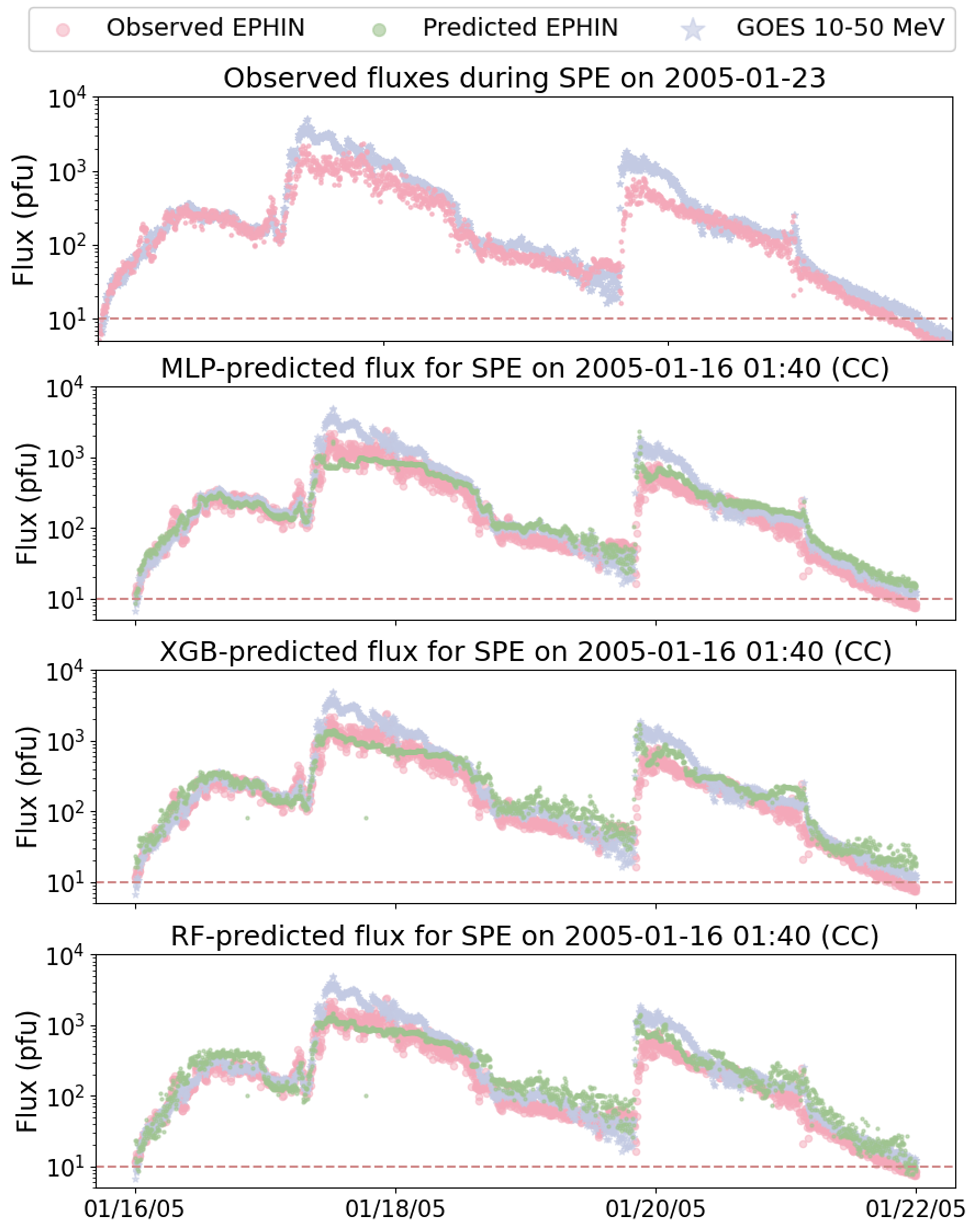} 
   \caption{Comparison of MLP, XGBoost, and RF flux predictions with fluxes observed by GOES and EPHIN during the SPE on 01/16/2005.}
 \label{comp}
\end{figure}

\begin{figure*}[ht!]
\centering 
\includegraphics[width=.6\textwidth]{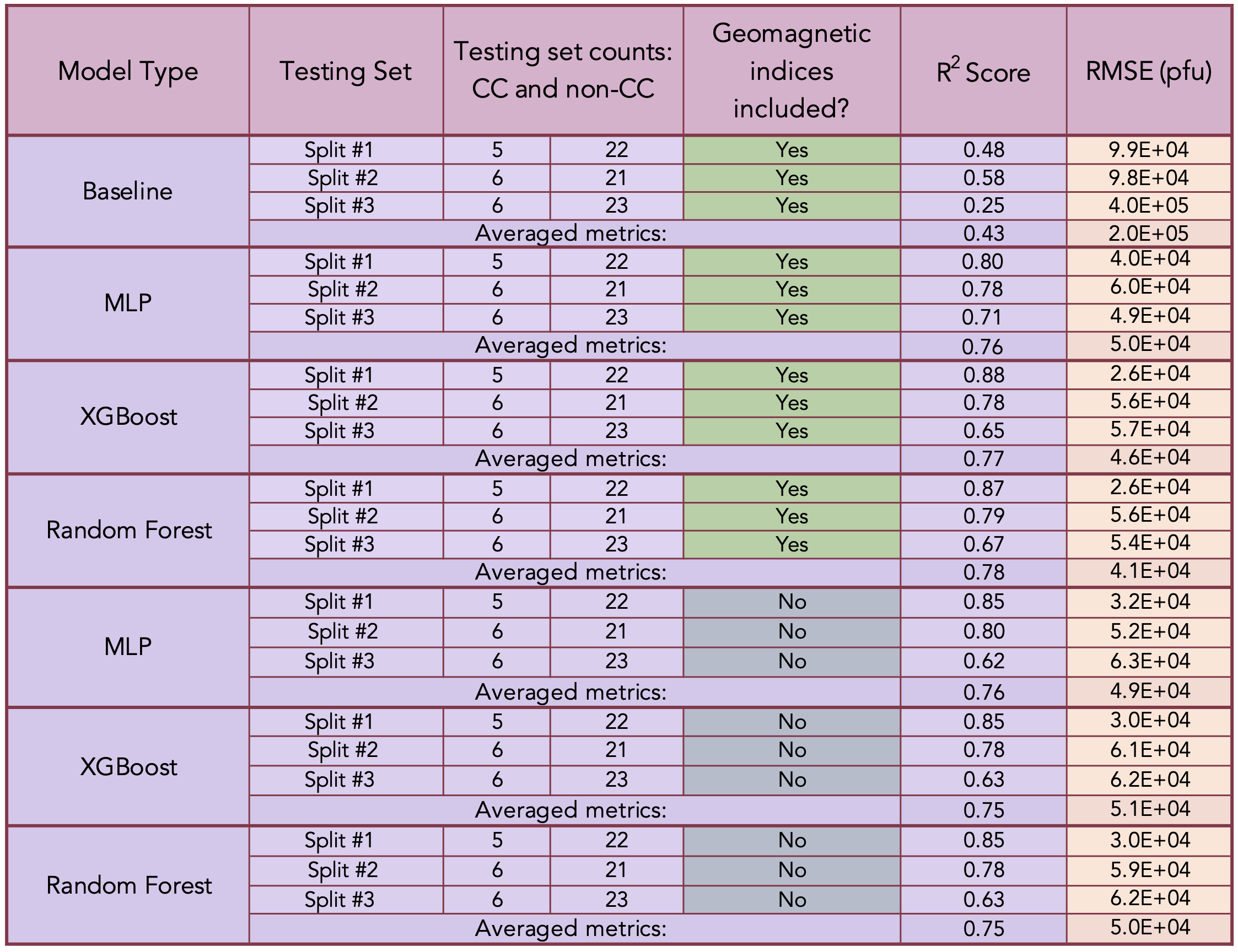} 
   \caption{Comparison of regression model performance in reconstructing `clean' SEP flux profiles from GOES data. Results are shown for three machine learning models (MLP, XGBoost, and RF) under two scenarios: with and without geomagnetic indices included as additional input features. For context, baseline results from a non-ML prediction method are also included.}
 \label{all_mods}
\end{figure*}

\begin{figure}[ht!]
\centering 
\includegraphics[width=.49\textwidth]{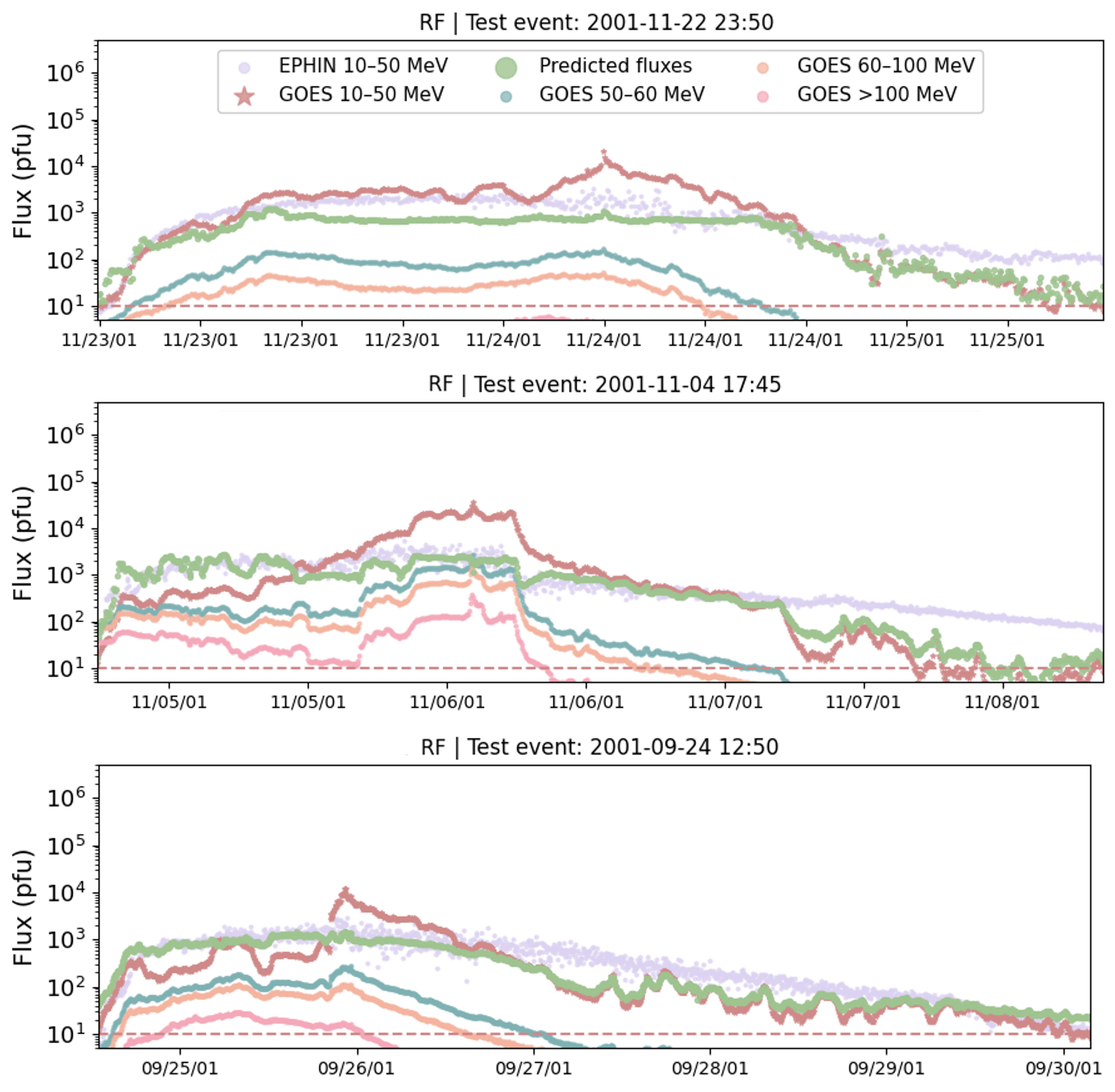} 
   \caption{True fluxes observed at L1 and GEO (in the 10–50 MeV, $\geq$50 -60 MeV, $\geq$60-100, and $\geq$100 MeV range) alongside RF-predicted fluxes for events classified as CCs. While higher-energy channels can significantly influence the observed profiles during these events, the model generally succeeds in disregarding contamination signatures and closely reproducing the true EPHIN fluxes.}\label{conts}
\end{figure}

To compare performance, all regression models are trained, validated, and tested using the data from the same SPEs (i.e., the train-validation-test splits were the same for each model). The input data consists of GOES proton fluxes across multiple energy channels (10 - 50\,MeV, $\geq$50\,MeV, $\geq$60\,MeV, and $\geq$100\,MeV) for a combination of CC \textit{and} non-CC events, alongside several geomagnetic indices (\textit{Ae, Kp, Ap} and \textit{Dst}). The target output in all cases is the corresponding EPHIN proton flux values, which serve as reference values representing uncontaminated proton flux profiles in this work. Figure \ref{comp} compares the true SEP fluxes observed at L1 and GEO with the model-predicted flux profiles for the same event across all three models. Among them, RF produces predictions that are most consistent with the target EPHIN measurements and demonstrates the best performance in capturing the onset of the event, which MLP and XGBoost consistently struggle to reproduce across most SPEs.

Table \ref{all_mods} summarizes the performance metrics obtained for each model configuration. The second column indicates which of the three data splits from our 83-event sample was used during the training, validation, and testing phases. Each split was constructed to be representative of the overall dataset, with comparable peak fluxes, fluences, and numbers of CCs. The third column specifies whether geomagnetic indices were included as additional input features alongside GOES proton fluxes across multiple energy channels. On average, the MLP achieves an $R^2$ of approximately 0.76 with an RMSE $\sim4.95\times10^4$ pfu$^{2}$, XGBoost also achieves $R^2$ $\sim$0.76 with an RMSE $\sim4.85\times10^4$ pfu$^{2}$, and RF attains $R^2$ $\sim$0.79 with an RMSE of $\sim4.55\times10^4$ pfu$^{2}$. Notably, the highest performance metrics for all models, both with and without geomagnetic indices, occur when Split \#1 is designated as the test set. Incorporating geomagnetic indices generally only slightly improves average $R^2$ scores and reduces RMSE values across most models, with RF showing the greatest improvement, increasing from an average $R^2$ of 0.75 to 0.83 and reducing RMSE from $5.0\times10^4$ to $4.1\times10^4$ pfu$^{2}$. MLP and XGBoost show more modest but consistent gains when these indices are included, suggesting that geomagnetic conditions provide valuable information for modeling contamination effects. With geomagnetic indices included, RF delivers the highest overall performance, followed by XGBoost, and then MLP. Without these indices, all three models perform comparably, with $R^2$ values $\sim$0.75 and RMSE values near $5.0\times10^4$ pfu$^{2}$. Performance variability between splits is evident, with Split \#3 serving as the test set consistently producing lower $R^2$ scores and higher RMSE values, indicating that certain events in this split, possibly those with unusual contamination patterns, are more challenging to model. Figure \ref{conts} shows RF-predicted fluxes for multiple CCs, and demonstrates how heavily-contaminated samples can be corrected using this model. These results highlight the benefit of incorporating geomagnetic data, particularly for tree-based models like RF, while also emphasizing that certain event subsets pose modeling challenges regardless of algorithm choice.

\section{Conclusions}
In this work, we tested data correction techniques to address flux contamination in GOES proton measurements using machine learning. The models used the GOES fluxes in the 10 - 50\,MeV, $\geq$50 - 60\,MeV, $\geq$50 - 60\,MeV, and $\geq$100\,MeV ranges, as well as multiple geomagnetic indices, and attempted to reconstruct SOHO/EPHIN 10 - 50\,MeV proton fluxes. Among three machine learning models (RF, XGBoost, and MLP) tested, we find RF performing most reliably, significantly reducing the contamination in GOES data (see Figure~\ref{conts} and Table~\ref{all_mods}) and demonstrating a principal possibility of implementing such an approach to GOES flux data correction. Continued refinement and optimization of such models may enhance the accuracy of SPE analysis using GOES data to improve the reliability of forecasting systems, and ultimately support the success of future space exploration missions.

\section*{Acknowledgments} 
We thank Bernd Heber and Arik Posner for their insights and suggestions on the EPHIN data analysis. This work was supported by the NSF FDSS grant 1936361, NASA LWS grant 80NSSC24K1111, and NASA HITS grant 80NSSC22K1561.

\bibliographystyle{IEEEtran}
 \bibliography{main}

@techreport{contamination_noaa,
  author = {Rodriguez, J. V.},
  title  = {EPEAD Electron Science Reprocessing Algorithm Theoretical Basis Document (ATBD)},
  institution = {National Centers for Environmental Information (NCEI), NOAA},
  year   = {2020},
  url    = {}
}

@article{goes_contamination_posner,
  author  = {Posner, A.},
  title   = {Up to 1-hour forecasting of radiation hazards from solar energetic ion events with relativistic electrons},
  journal = {Space Weather},
  volume  = {5},
  number  = {5},
  pages   = {05001},
  month   = may,
  year    = {2007},
  doi     = {10.1029/2006SW000268}
}

@article{rodriguez_cont,
  author  = {Rodriguez, J. V. and others},
  title   = {Validation of the effect of cross-calibrated GOES solar proton effective energies on derived integral fluxes by comparison with STEREO observations},
  journal = {Space Weather},
  volume  = {15},
  number  = {2},
  pages   = {290--309},
  month   = feb,
  year    = {2017},
  doi     = {10.1002/2016SW001533}
}

@article{sep_energies_anastasiadis,
  author  = {Anastasiadis, A. and others},
  title   = {Solar energetic particles in the inner heliosphere: status and open questions},
  journal = {Philosophical Transactions of the Royal Society A},
  volume  = {377},
  number  = {2148},
  pages   = {20180100},
  month   = jul,
  year    = {2019},
  doi     = {10.1098/rsta.2018.0100}
}

@article{health_risk_onorato,
  author  = {Onorato, G. and others},
  title   = {Understanding the Effects of Deep Space Radiation on Nervous System: The Role of Genetically Tractable Experimental Models},
  journal = {Frontiers in Physics},
  volume  = {8},
  pages   = {362},
  year    = {2020},
  doi     = {10.3389/fphy.2020.00362}
}

@article{sickness_lee,
  author  = {Lee, C. O. and others},
  title   = {Observations and impacts of the 10 September 2017 solar events at Mars: An overview and synthesis of the initial results},
  journal = {Geophysical Research Letters},
  volume  = {45},
  number  = {17},
  pages   = {8871--8885},
  month   = sep,
  year    = {2018},
  doi     = {10.1029/2018GL079162}
}

@article{radiation_dandouras,
  author  = {Dandouras, I. and Roussos, E.},
  title   = {High-energy particle observations from the Moon},
  journal = {Philosophical Transactions of the Royal Society A},
  volume  = {382},
  number  = {2251},
  pages   = {20230311},
  year    = {2024},
  doi     = {10.1098/rsta.2023.0311}
}

@article{bsphere_deflecion_liu,
  author  = {Liu, J. and others},
  title   = {Solar flare effects in the Earth's magnetosphere},
  journal = {Nature Physics},
  volume  = {17},
  number  = {7},
  pages   = {807--812},
  year    = {2021},
  doi     = {10.1038/s41567-021-01203-5}
}

@article{main_moon_liuzzo,
  author  = {Liuzzo, L. and Poppe, A. R. and Lee, C. O. and Xu, S. and Angelopoulos, V.},
  title   = {Unrestricted solar energetic particle access to the Moon while within the terrestrial magnetotail},
  journal = {Geophysical Research Letters},
  volume  = {50},
  pages   = {e2023GL103990},
  year    = {2023},
  doi     = {10.1029/2023GL103990}
}

@article{me1,
  author  = {Ali, A. and others},
  title   = {Predicting Solar Proton Events of Solar Cycles 22--24 Using GOES Proton and Soft-X-Ray Flux Features},
  journal = {Astrophysical Journal Supplement Series},
  volume  = {270},
  number  = {1},
  pages   = {15},
  month   = jan,
  year    = {2024},
  doi     = {10.3847/1538-4365/ad0a6c}
}

@unpublished{sep_comparison_ali,
  author  = {Ali, A. and Sadykov, V.},
  title   = {Comparative Analysis of 10--50 MeV Solar Proton Events at Lagrange Point 1 and the Geostationary Orbit},
  note    = {in preparation},
  year    = {2025}
}

@article{transport_effects_battarbee,
  author  = {Battarbee, M. and others},
  title   = {Multi-spacecraft observations and transport simulations of solar energetic particles for the May 17th 2012 event},
  journal = {Astronomy \& Astrophysics},
  volume  = {612},
  pages   = {A116},
  month   = may,
  year    = {2018},
  doi     = {10.1051/0004-6361/201731451}
}

@article{pl_example_aminalragia,
  author  = {Aminalragia-Giamini, S. and others},
  title   = {Solar Energetic Particle Event occurrence prediction using Solar Flare Soft X-ray measurements and Machine Learning},
  journal = {Journal of Space Weather and Space Climate},
  volume  = {11},
  pages   = {59},
  month   = nov,
  year    = {2021},
  doi     = {10.1051/swsc/2021043}
}

@article{particle_energy_obrien,
  author  = {O'Brien, T. P. and Mazur, J. E. and Looper, M. D.},
  title   = {Solar Energetic Proton Access to the Magnetosphere During the 10--14 September 2017 Particle Event},
  journal = {Space Weather},
  volume  = {16},
  number  = {12},
  pages   = {2022--2037},
  month   = dec,
  year    = {2018},
  doi     = {10.1029/2018SW001960}
}

@article{arrival_direction_filwett,
  author  = {Filwett, R. J. and Jaynes, A. N. and Baker, D. N. and Kanekal, S. G. and Kress, B. and Blake, J. B.},
  title   = {Solar Energetic Proton Access to the Near-Equatorial Inner Magnetosphere},
  journal = {Journal of Geophysical Research: Space Physics},
  volume  = {125},
  number  = {6},
  pages   = {e2019JA027584},
  month   = jun,
  year    = {2020},
  doi     = {10.1029/2019JA027584}
}

@article{kress_cutoff,
  author  = {Kress, B. T. and Hudson, M. K. and Selesnick, R. S. and Mertens, C. J. and Engel, M.},
  title   = {Modeling geomagnetic cutoffs for space weather applications},
  journal = {Journal of Geophysical Research: Space Physics},
  volume  = {120},
  number  = {7},
  pages   = {5694--5702},
  month   = jul,
  year    = {2015},
  doi     = {10.1002/2014JA020899}
}

@misc{stumpo_rf,
  author = {Stumpo, G. and Giordano, S. and Laurenza, M. and Cliver, E. W.},
  title  = {Random forest regression model for predicting the onset and intensity of solar proton events using relativistic electron data},
  howpublished = {arXiv preprint},
  eprint = {2406.12730},
  archivePrefix = {arXiv},
  month  = jun,
  year   = {2024},
  url    = {https://arxiv.org/pdf/2406.12730}
}

@article{leemyr,
  author  = {Miller, D. A. B. and Weimer, D. R. and Anderson, B. J.},
  title   = {LEEMYR: A machine learning model for predicting low-energy electron flux at geostationary orbit},
  journal = {Space Weather},
  volume  = {22},
  number  = {5},
  pages   = {e2024SW003962},
  month   = may,
  year    = {2024},
  doi     = {10.1029/2024SW003962}
}

@article{bailey,
  author  = {Bailey, R. L. and Reiss, M. A. and Arge, C. N. and M{\"o}stl, C. and Owens, M. J. and Amerstorfer, U. V. and Henney, C. J. and Amerstorfer, T. and Weiss, A. J. and Hinterreiter, J.},
  title   = {Using gradient boosting regression to improve ambient solar wind model predictions},
  journal = {Space Weather},
  volume  = {19},
  number  = {5},
  pages   = {e02673},
  month   = may,
  year    = {2021},
  doi     = {10.1029/2020SW002673}
}

\vspace{12pt}
\end{document}